\documentclass[runningheads]{llncs}
\usepackage[T1]{fontenc}
\usepackage[utf8]{inputenc}
\usepackage{gensymb}
\usepackage{cite}
\usepackage{graphicx}
\usepackage{tabularx}
\usepackage{booktabs}

\begin{document}
\title{RESISTO Project: Safeguarding the Power Grid from Meteorological Phenomena}
\titlerunning{RESISTO Project}
%

\author{Jacob Rodríguez-Rivero \inst{1} \and David López-García \inst{2} \and Fermín Segovia \inst{2} \and Javier Ramírez \inst{2} \and Juan Manuel Górriz \inst{2} \and R. Serrano \inst{3} \and D. Pérez \inst{3} \and Ivan Maza \inst{4} \and Anibal Ollero \inst{4} \and Pol Paradell Solà \inst{5} \and Albert Gili Selga \inst{5} \and Jose Luis Domínguez-García \inst{5} \and A. Romero \inst{1} \and A. Berro \inst{1} \and Rocío Domínguez \inst{1}  \and Inmaculada Prieto \inst{1}}
\authorrunning{Jacob Rodríguez-Rivero et al.}
%
\institute{e-Distribución Redes Digitales, Spain \\
\and Department of Signal Theory, Networking and Communications, University of Granada, Spain \\
\and ATIS Solutions, Spain
\and GRVC Robotics Lab, University of Seville, Spain \\
\and Catalonia Institut for Energy Research---IREC, Power Electronics Department; Jardins de les Dones de Negre 1, 2ª pl., Sant Adrià del Besòs, 08930 Barcelona, Spain\\} 
\maketitle              
\begin{abstract}
The RESISTO project, a pioneer innovation initiative in Europe, endeavors to enhance the resilience of electrical networks against extreme weather events and associated risks. Emphasizing intelligence and flexibility within distribution networks, RESISTO aims to address climatic and physical incidents comprehensively, fostering resilience across planning, response, recovery, and adaptation phases. Leveraging advanced technologies including AI, IoT sensors, and aerial robots, RESISTO integrates prediction, detection, and mitigation strategies to optimize network operation. This article summarizes the main technical aspects of the proposed solutions to meet the aforementioned objectives, including the development of a climate risk detection platform, an IoT-based monitoring and anomaly detection network, and a fleet of intelligent aerial robots. Each contributing to the project's overarching objectives of enhancing network resilience and operational efficiency.

\end{abstract}
\section{Introduction}

The RESISTO project aims to enhance the resilience of electrical networks, minimizing the impact of extreme climate events and related risks on service provision. In response to the increasing frequency of such events, as highlighted in the latest IPCC report \cite{IntergovernmentalPanelonClimateChangeIPCC2023}, RESISTO emphasizes intelligence and flexibility within distribution networks to effectively mitigate these challenges. The project adopts a comprehensive approach to resilience, addressing climatic events (water, wind, fire) as well as physical incidents (human, animal, others). Innovations and developments across all resilience conceptual phases (Planning, Response, Recovery, and Adaptation) are central to RESISTO's objectives.

\begin{itemize}

\item In the planning phase, RESISTO focuses on prediction and prevention by leveraging existing tools to identify high-risk areas under specific climatic conditions and utilizing climate prediction tools. Moreover, the project assesses potential risks from physical actions, including those from animals and humans, with the consideration of aerial robots to gather data and images for integration into planned models.

\item During the response phase, RESISTO emphasizes detection, proposing options for coarse sensing through real-time grid monitoring and nearby IoT-based sensors. Aerial robots play a crucial role in minimizing detection time for overhead line failures, verifying warnings, and assessing environmental conditions. Advanced 3D systems aid in creating virtual decision tables for response and mitigation actions.

\item In the recovery/mitigation phase, RESISTO explores several strategies, including the use of Dynamic Thermal Line Rating (DTLR) to optimize power flows and integrate with grid reconfiguration algorithms. Additionally, RESISTO explores the role of renewables in enhancing network stability and security, offering functionalities such as ``Black-Start'' and ``Grid-Forming''.

\item In the adaptation phase, RESISTO focuses on learning to improve resilience, incorporating aerial robots to generate data for integration into machine learning methods. This phase also involves equipment planning and adaptation through the use of 3D digital twins of electrical networks, facilitating visualization and understanding of decisions and actions' impacts.

\end{itemize}

\subsection{Objectives}

The primary technical objectives of the study are interconnected and aimed at addressing the previously mentioned challenge.
\begin{itemize}

\item Firstly, a risk assessment tool with diverse time horizons is proposed to identify potential risks in the electrical network, enabling the implementation of detection and early warning systems for short-term (days) and long-term (years) risk mitigation strategies. 

\item Secondly, the development and application of technology are emphasized to enhance the monitoring and control capacity of the network, transforming it into both an energy supply structure and a data interaction hub for environmental risk information dissemination and proactive response.

\item Thirdly, optimization of existing infrastructure and reduction of investments in new network infrastructure are targeted through the development of advanced control and management systems. 

\item Fourthly, digitalization and the utilization of IoT sensors, coupled with aerial robots or drones, are advocated for measuring environmental variables and validating alerts, thereby minimizing personal risk and response time. Additionally, AI-powered algorithms for efficient risk management and 3D visualization systems for network control identification are proposed. 

\item Finally, the study aims to demonstrate and validate these developments through prototype implementation in diverse pilot areas, highlighting scalability and replicability for broad national and international distribution network impact.

\end{itemize}

\subsection{Project innovations:}
The project introduces several innovations aimed at maximizing and ensuring the achievement of the previously established objectives. These innovations are outlined as follows:

\begin{itemize}

\item \textbf{Innovation 1 - Climate Risk Detection Platform:}
The project proposes the development of a platform dedicated to detecting climate risks, vulnerabilities, load optimization, and providing recommendations or reconfigurations for decision-making. This platform will leverage machine learning techniques to enhance its capabilities.
\item \textbf{Innovation 2 - IoT-Based Monitoring Network:}
Another key innovation involves the establishment of an Internet of Things (IoT)-based monitoring network comprising gateways, sensors, weather stations, thermal cameras, and tele-surveillance cameras. These components will be strategically deployed to gather real-time data, which will then be integrated into a unified risk platform for comprehensive analysis.
\item \textbf{Innovation 3 - Fleet of Intelligent Aerial Robots:}
The project proposes the deployment of a fleet of intelligent aerial robots designed to autonomously validate warnings. This innovative approach aims to minimize risks for operators while simultaneously reducing response time in emergency situations. The integration of aerial robots into the system enhances the project's capacity for swift and efficient risk assessment and management.

\end{itemize}

\section{Proposed solution}
This section outlines three key components designed to bolster the resilience of electrical grids. The first is the GridWatch Electrical Resilience Platform, a creation of IREC, which amalgamates climate data with the electrical transportation and distribution system. This platform centralizes all relevant information for the visualization of electrical system interactions, integrating external diagnostic tools and employing mathematical algorithms and AI for fault location and resolution. It incorporates various tools such as local weather stations, online weather systems, fire sensors, infrared cameras, and a fleet of drones. The second component focuses on the Automatic Detection of Operation Temperature Anomalies using Thermal Imaging, emphasizing the monitoring of power transformer temperatures through thermal imaging technology. It introduces a system architecture deployed by the University of Granada and ATIS with 20 thermal cameras, 9 industrial PCs, and a continuous anomaly detection system based on segmentation and an adaptive prediction model. The third component, developed by members of the University of Seville, addresses the limitations of current drone applications in electrical grid inspection and introduces a multi-drone system, combining multi-rotors and fixed-wing VTOLs for efficient power line inspection. The overall goal of these components is to augment the resilience and efficiency of electrical grids through advanced monitoring, fault detection, and inspection facilitated by a blend of technological tools, algorithms, and drone systems.

\subsection{Electrical resilience platform: GridWatch}
\begin{figure}
\includegraphics[scale=0.062]
{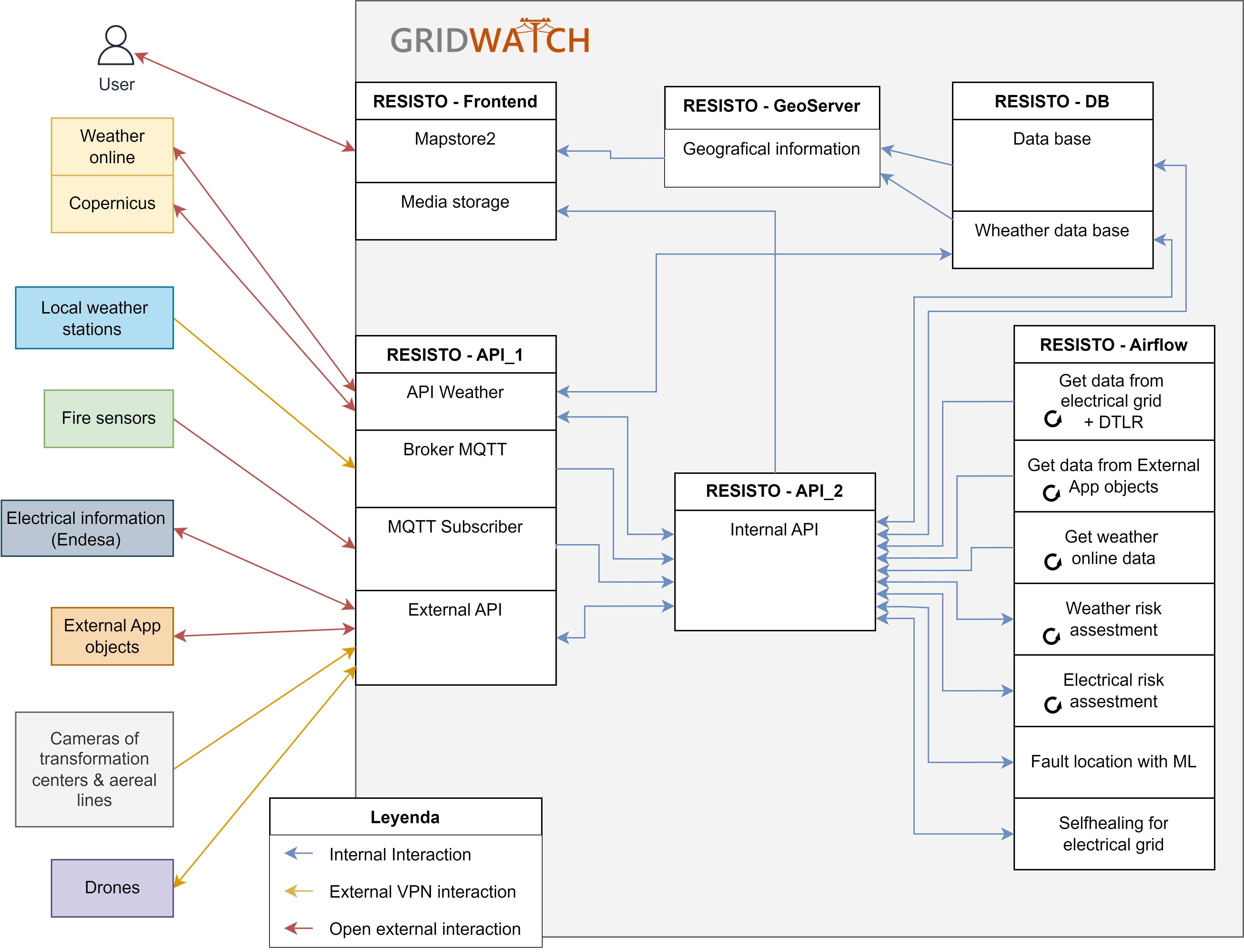}
\caption{GridWatch platform components and external interactions.} \label{GridWatch_interactions}
\end{figure}

The GridWatch electrical resilience platform, developed by IREC, has been developed to unite climate information with the electrical transportation and distribution system. Figure \ref{GridWatch_interactions} describes the principal elements of the GridWatch platform and the external interaction with the other elements involved in the project. This platform mainly fulfils three objectives:

\begin{enumerate}
  \item Concentrate all the information in a single place where the interactions of the electrical system can be visualised.
  \item Integrate external diagnostic and review tools to view and verify incidents.
  \item Use different mathematical algorithms or artificial intelligence to locate electrical faults and how to solve them.
\end{enumerate}

Mapstore2\cite{mapstore_web}, an open-source tool for data visualisation, has been used for the platform visualisation system. Geoserver\cite{geoserver_web}, a geospatial information publishing tool, has also been used to render all the information on the platform.

Different tools have been integrated into the platform to have maximum information and thus be able to make the best decisions. The most notable systems integrated with the platform are detailed below:

\begin{itemize}
  \item Local weather stations. A dozen stations have been installed in the vicinity of the pilot using Lora communications and mobile communications to send information to servers that, in turn, use the MQTT protocol to communicate measurements.
  \item Online weather system. Different sources, such as AEMET and Copernicus, have been used to obtain information on the climatological state.
  \item Fire sensors. Sensors capable of detecting forest fires have been integrated. These sensors send the information through MQTT.
  \item infrared cameras in transformation centres. These send possible anomaly detection alerts on the platform to be visualised there.
  \item The platform has been integrated with a fleet of drones. In this case, the platform can send the coordinates to be reviewed by the drones and receive the results of this inspection.
  \item Integration with the electrical system. The platform has integrated the available information from the electrical network, and for cases in which there is no observability, a Power Flow system has been implemented to determine the state of the network using the mathematical model modelled during this project.
\end{itemize}

On the other hand, the platform integrates different algorithms to increase the resilience of the network:

\begin{itemize}
    \item Algorithm of effects on the electrical network. This algorithm calculates the possible interference of weather events with overhead power lines, such as temperature, wind, or fire risks.
    \item Recalculates the capacity of the lines. The DTLR algorithm is capable of recalculating the maximum capacity of the lines with the weather conditions; in this way, with more information on the status of these lines, better decisions can be made to redirect electricity or find critical points.
    \item Fault location system in the electrical system\cite{ml_fault_location}. This algorithm developed at IREC uses machine learning methods to locate the location of the electrical fault based on the state of the last network cycles before the fault and the time of the fault. This method reduces the time to locate electrical faults since these systems are previously trained.
    \item Network reconfiguration system\cite{islanding_algorithm}. The platform also has an algorithm that, once the failure occurs, proposes new network configurations to keep the maximum number of clients connected. This algorithm uses a heuristic method to find mathematical solutions; in this case, it uses a genetic algorithm to determine the best solution in a short time.
\end{itemize}

\subsection{Automatic detection of operation temperature anomalies using thermal imaging}

The oversight of electricity networks requires meticulous attention to potential overloads \cite{Vitolina2015,Mullerova2015}, given their significant impact on power transformers, responsible for approximately 12\% of total failures \cite{Yazdani-Asrami2015}. Transformer malfunctions, in particular, present a significant hazard, capable of igniting fires with potentially severe economic and personal consequences due to the urban positioning of transformers \cite{dolata2016online}. To mitigate these failures, routine maintenance of installations is essential, including regular measurements. These measurements, among others \cite{Bakar2014}, involve monitoring operational temperatures \cite{Kunicki2020}. Consequently, measuring both transformer and ambient temperatures is a fundamental component of standard protocols for power transformer monitoring \cite{Peimankar2017,Velasquez-Contreras2011}. These parameters are pivotal variables requiring accurate and uninterrupted measurement and can also serve as predictors for other variables, such as reactive power or current intensity \cite{ramirez2020power}.

Temperature measurement of power transformers conventionally relies on integrated sensors, albeit susceptible to potential issues, limiting predictive capabilities for malfunctions \cite{DeMelo2021}. In contrast, thermal imaging, a well-established technology offers advantages including independence from transformers, scalability, and diverse application effectiveness \cite{Sirca2018, Lu2021, bagavathiappan2008condition, Itami2004}. It operates without direct physical contact, minimizing susceptibility to transformer failures such as temperature peaks or fluctuations that may harm sensors. This technology not only monitors transformer temperature comprehensively but also individual components or adjacent equipment independently. Recent proposals suggest its use for correcting and monitoring temperature-related issues in electrical substations and power transformers \cite{Zarco-Perinan2021, Segovia2023, martinez2019prediction}. 
Early failure prevention in the electrical grid and risk reduction are closely tied to artificial intelligence methods enabling the identification and notification of potential failures. The widespread impact of artificial intelligence, particularly big data and machine learning, extends across various societal domains\cite{Gorriz2020, Gorriz2023}, including autonomous transportation \cite{Ma2020}, neuroscience \cite{Lopez-Garcia2022, Lopez-Garcia2022a, Jafari2023a}, healthcare\cite{Lopez-Garcia2018,Jafari2023} or automatic text translation \cite{Mohamed2024}. Notably, the management of electrical grids has been significantly influenced by artificial intelligence advancements, empowering grids to make informed decisions in response to shifts in energy demand, renewable energy production, and extreme weather events. Furthermore, the escalating volume of data enables the application of machine learning for detecting and preventing anomalous behavior of the electrical network.

\begin{figure}
\includegraphics[width=\textwidth]{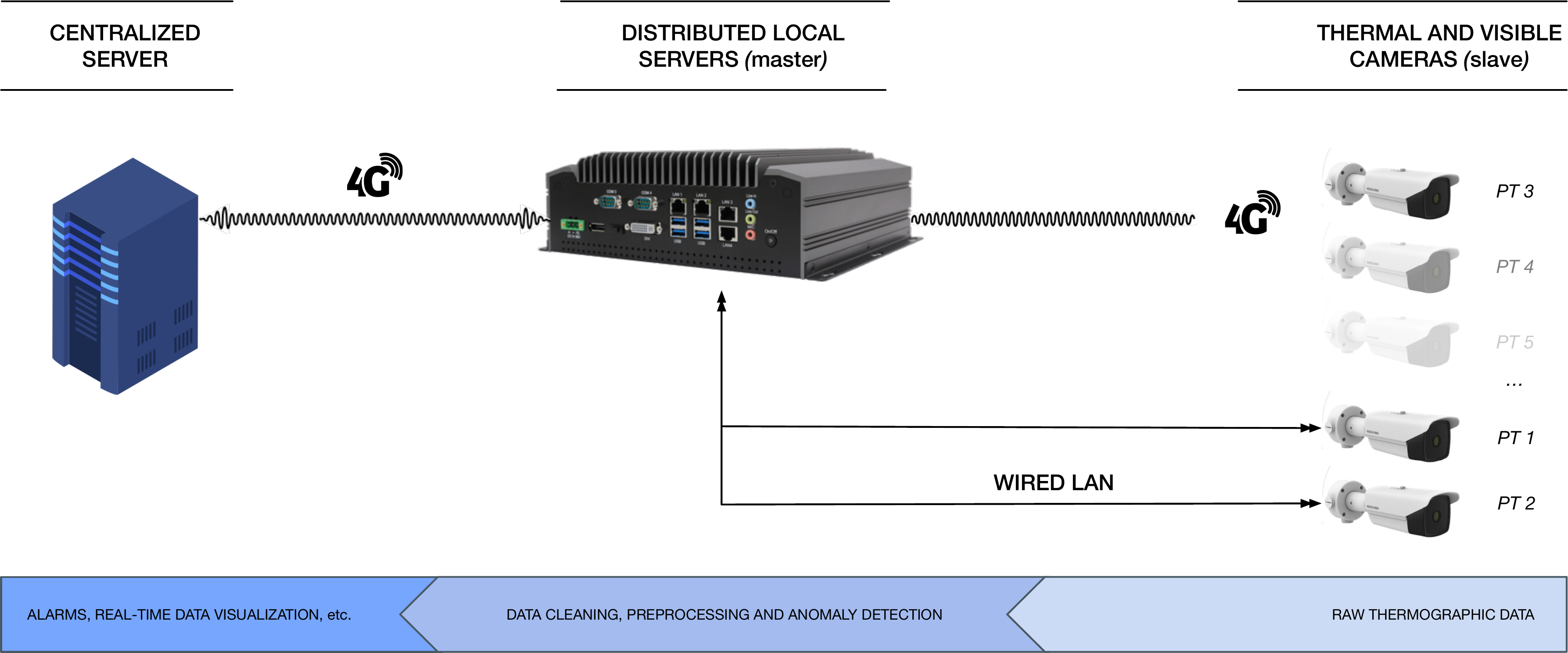}
\caption{Thermographic data acquisition architecture.} \label{architecture}
\end{figure}

\subsubsection{System architecture} The proposed thermography data acquisition and processing system consists of 20 thermal cameras and 9 industrial PCs for control, as depicted in Figure \ref{architecture}. Industrial PCs establish connections with cameras either via physical Ethernet cables when co-located or through 4G communication when at different sites. To ensure security, radio-link connections are VPN-secured, although their use was dismissed due to inadequate coverage in some locations.

Thermal image capture is initiated by industrial PCs, with acquisition frequency determined by connection type (one image per minute for physical links, one every 5 minutes for radio links). Upon receipt, industrial PCs handle storage and processing tasks. Additionally, industrial PCs are equipped with 4G connectivity for tasks such as monitoring, control, and data extraction, with external access facilitated through the SSH protocol, all secured by VPN for enhanced security.

\subsubsection{Thermal anomalies detection system.}
\begin{figure}
\includegraphics[width=\textwidth]{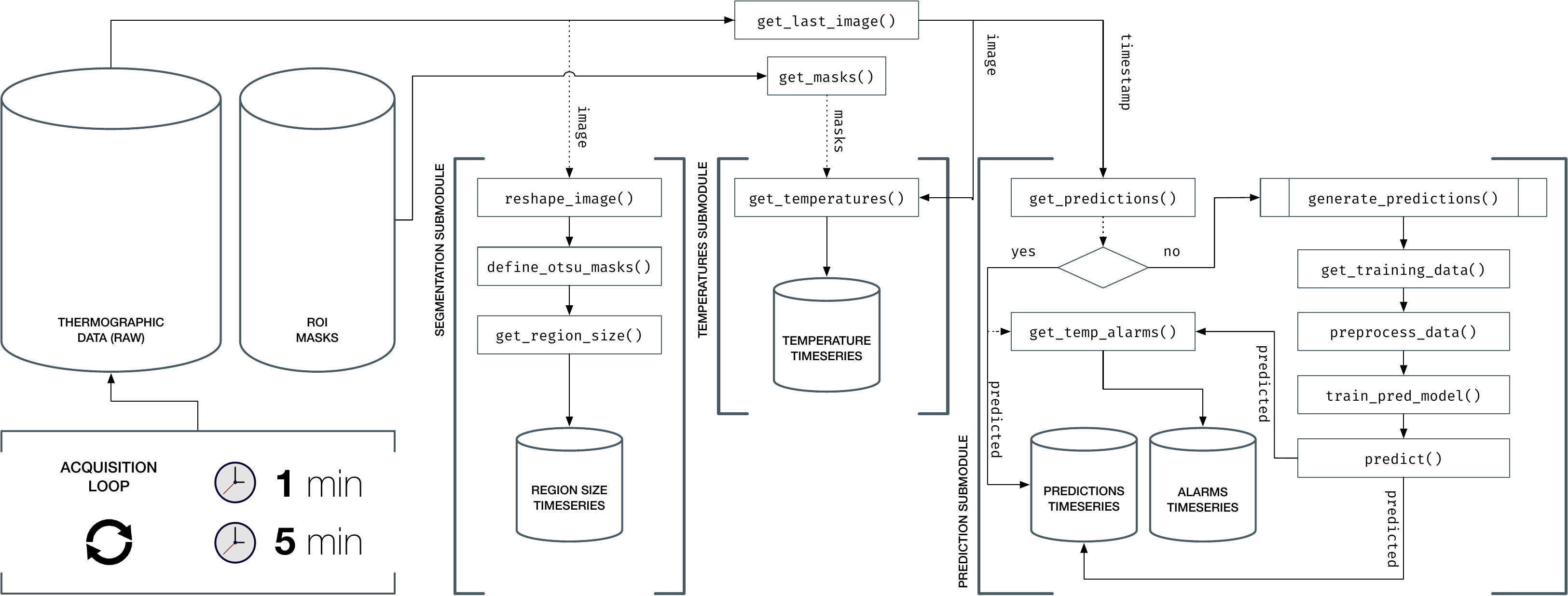}
\caption{Thermal anomalies detection system diagram.} \label{diagram}
\end{figure}

The proposed thermal anomaly detection solution operates continuously on distributed servers independently of the thermal image acquisition system. Upon receiving a captured thermal image, the server processes it to check if the registered temperatures of transformer regions fall within optimal margins predicted by an adaptive model; if not, a thermal anomaly alarm is triggered. The system's flowchart is illustrated in Figure \ref{diagram}, and a detailed description of each module follows.

\begin{itemize}
\item The thermographic data processing involves segmentation and temperature extraction submodules. Segmentation occurs through automated algorithms and manually-defined masks. Nine regions of interest (ROIs) were manually defined \ref{camera-masks}, including primary and secondary windings terminals, high and low voltage bushings, transformer body, and background temperature. Individual masks are crafted for each camera scene, allowing the computation of the mean temperature of the top 5\% of pixels within each ROI. Automatic segmentation is computed employing Otsu \cite{Otsu1996} and Maximally Stable Extremal Regions Algorithm (MSER) \cite{Matas2004} algorithms, to store the size of each ROI and identify deviations indicating potential anomalies such as vegetation growth or intruders.

\begin{figure}
\includegraphics[width=\textwidth]{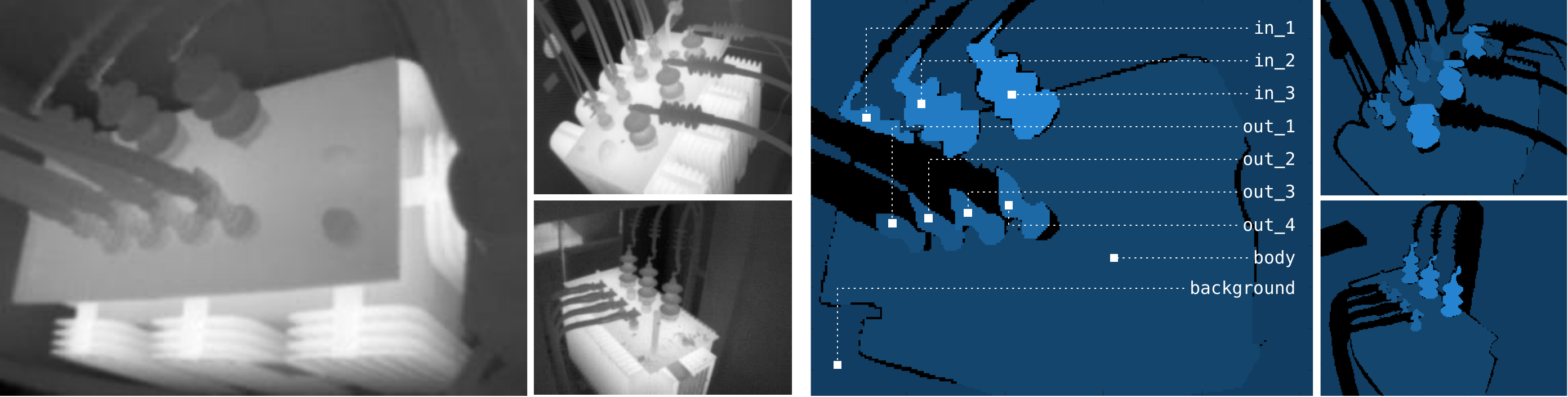}
\caption{Raw thermal data and manually-defined regions of interest.} \label{camera-masks}
\end{figure}

\item The prediction model and alarms submodule plays a critical role in evaluating thermal data extracted by the segmentation and temperature extraction submodules. Following image capture and processing, this submodule assesses if the extracted temperatures meet expected margins for each region. If anomalies are detected, specific region alarms are generated and relayed to the central server for operator response. These margins are dynamic, contingent upon factors such as time of day, ambient temperature, season, weather conditions, and transformer workload. Notably, the temperature data captured by the cameras exhibits stationarity over time, resulting in daily "duck curves" representing the transformer's load profile across 24 hours. Leveraging this stationarity, future predictions are adaptively calculated based on historical observations. The proposed adaptive prediction model undergoes recurrent updates based on newly available temperature data, a characteristic known as online learning, which is particularly advantageous in dynamic environments with rapidly changing training data. However, determining the retraining frequency is crucial to balance prediction accuracy with computational efficiency and the risk of overfitting. In the proposed solution, the adaptive prediction model is retrained every 720 minutes (12 hours), aligning with the frequency of thermal image acquisition. This approach ensures that the model remains responsive to the transformer's evolving temperature conditions while avoiding excessive computational costs. By incorporating the latest 24 hours of temperature data into the training process, the model adapts to the current temperature dynamics, enabling accurate predictions for the subsequent 12-hour time window. Thus, the model's adaptability enhances its performance in forecasting temperature variations. The prediction and alarm generation submodule verifies then the difference between the recorded temperature and the estimated value against a predefined threshold (set at a margin of 15 degrees C) for each region of interest. This threshold can be adjusted to fine-tune the sensitivity of the anomaly detection system. Subsequently, a temporary table is generated, encompassing the recorded temperature data, the predicted temperature values, and a control bit associated with each ROI indicating the presence or absence of a thermal anomaly alarm. This information is then extracted and transmitted to the central server via an accessible API.
\end{itemize}

\subsection{Fleet of Drones}

Nowadays, multi-rotors, with small range and endurance, and usually with remote pilots in the visual line of sight, are applied for local inspections of transmission towers~\cite{he2019research,baik2019unmanned} or grid segments~\cite{iversen2021pneumatic}. However, this is not enough for long-endurance inspection of the electrical grid composed of medium and high-voltage lines. Motivated by this, new research lines are arising to mitigate the limitations of these robots~\cite{iversen2021design}. There are also long range and endurance UAVs, usually large and heavy, which are used in military  application, surveillance, maritime applications, environmental surveillance, and others. In addition, small fixed-wing systems, usually flying below 150 meters,  are also applied  for surveillance and mapping of relatively large areas. However, they have constraints related to the accuracy due to the limitations of the on-board sensing systems. In the last years, new systems with vertical take-off and landing and flight as fixed wing have been developed. Different types of  systems exist, including those with different propellers to fly as fixed wing and vertical take-off and landing (VTOL), as well as systems that can change the orientation of the propellers after take-off to fly as fixed wing and later changing again to the initial configuration for landing. However, these platforms are currently not efficient to perform the hovering required for detailed very close view and aerial manipulation for maintenance operation.

The inspection of large infrastructures such as electrical grid systems can be performed more efficiently with multiple UAVs~\cite{deng2014unmanned}. The advantages when comparing with the use of a single UAV are: decreasing the total time for the whole inspection; minimizing delays for event detection; application of teaming techniques involving multiple specialized platforms for example to obtain closed view in local inspection and have long-range inspection; and improved reliability by avoiding dependencies of a single UAV.

The drones chosen for the power lines inspection are heterogeneous since the system includes multi-rotors and the fixed-wing VTOLs. This configuration of the team covers long range operation along the power lines, but also a detailed and close inspection of the transmission towers, including their grounding system in the base. However, as the flight time of the aerial robots is nowadays limited, compared to the large size of the power lines, the system also includes battery recharging stations where the UAVs autonomously land and take-off to extend their range and perform continuous operation.

\begin{figure}[htbp!]
\centering
\includegraphics[width=0.5\textwidth]{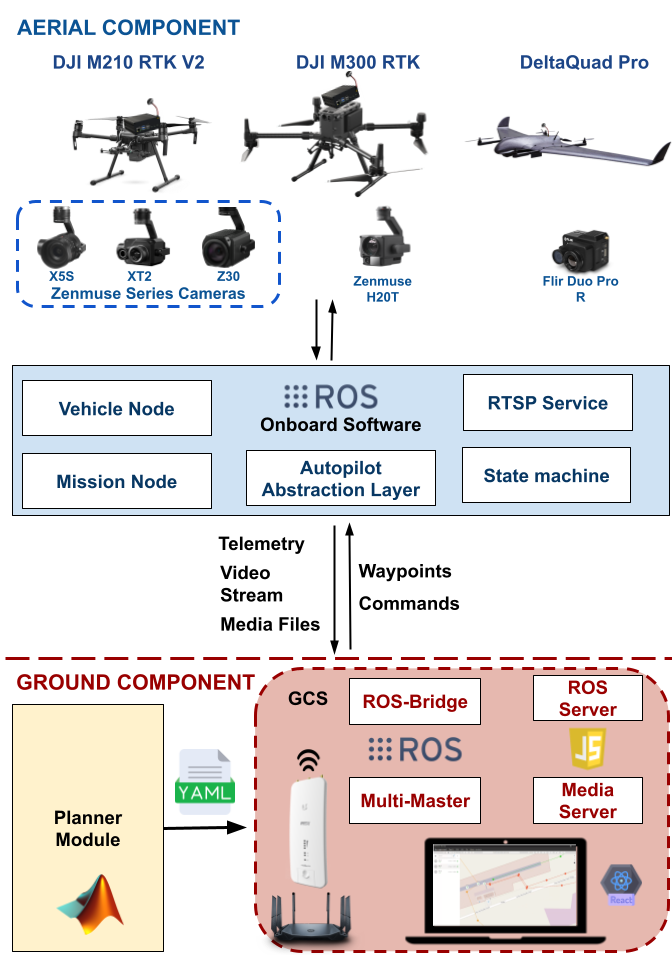}
\caption{Multi-drone system architecture with the ground segment composed by the Ground Control Station (GCS) and the mission planner module at the bottom, the main software modules on-board the drones in the middle and the drones and their payloads on top.} 
    \label{fig:drone-arch}
\end{figure}

A classification of different possible architectures for multi-drone systems can be found in~\cite{maza_arch_UAVhandbook15}, and the scheme that can fit better for the application addressed in this project is an architecture for intentional cooperation. An example of this type of architecture applied to a multi-drone team in field tests can be found in~\cite{maza_jfr11_multiuav} and we have adopted that approach in RESISTO. The architecture is endowed with different modules that solve the usual problems that arise during the execution of multipurpose missions, such as task allocation, conflict resolution, task decomposition, and sensor data fusion. The approach had to satisfy two main requirements: robustness for operation in power line inspection scenarios and easy integration of different autonomous vehicles. An overview of the architecture of the multi-drone system is shown in Figure~\ref{fig:drone-arch} with the drones and their payloads on top, the main software modules on-board the drones are in the middle and the Ground Control Station (GCS) and the mission planner are at the bottom. The software development is mainly based on the Robot Operating System (ROS)~\cite{ROS} which stands as the established standard for developing robot applications, extending its utility to the domain of autonomous vehicles such as drones. Interacting directly with the autopilot at the lowest level, we have developed several ROS nodes that makes possible to control drones with autopilots supporting the MAVLink protocol, and or DJI autopilots. In addition, SMACH~\cite{smach} is the ROS library applied for monitoring the mission inside the drones and the GCS.

The mission planner module autonomously computes the flight plans for all the drones. The tasks and all their associated parameters for each drone are defined in a file in YAML format. This YAML file plays a pivotal role in configuring precise flight paths and parameters for each UAV (such as the gimbal and heading angle, camera triggers for videos or pictures, speed, etc.) ensuring an efficient inspection.

To facilitate comprehensive mission management and supervision, an user-friendly graphical interface using React in JavaScript was developed for the GCS. This interface is essential for efficiently handling mission tasks, including loading, editing, commanding and monitoring.

Regarding the ROS nodes running within the on-board computer of the drone, the primary one is the vehicle node, which handles the drone's state and movement collecting data from the on-board sensors, and allows the interaction with the GCS thanks to the ROS communication options. Its functions are acting as a server to the client that will be at the GCS broadcasting telemetry data and providing basic control commands services. 

The node that takes advantage of this vehicle node is the mission node, which will be an abstraction layer with higher level functionalities related to the mission. The whole mission process from its upload to its end is governed by a state machine, which monitors the connection with the GCS, correct upload and command actions and the running process gathering information and making this information transparent regardless of the autopilot and useful to visualize it with the Graphical User Interface.


The communication link between the Ground and Aerial components relies on the ROS multimaster-fkie~\cite{multimaster} tool. This package enables the presence of a ROS-Master node on each CPU, facilitating the sharing of services and topics across a common network. Notably, the ROS nodes operating on the vehicles remain independent of the Ground Component once the mission starts, ensuring their continuous operation even if the connection to the GCS is lost. 

Finally, a node to stream the video from the on-board cameras has been developed to operate through different streaming paradigms. The drones chosen for our application are the DeltaQuad Pro as the VTOL fixed-wing and the DJI Matrice Series drones as the multicopter part (DJI Matrice 210 V2 RTK and DJI Matrice 300 RTK). All the drones are equipped with both visual and thermal cameras on-board.

\section{Discussion}
To align with the overarching goals of the project, specific objectives have been delineated, focusing on the development of technology, supported by both hardware and software, essential for fulfilling the aforementioned general objectives. The scientific and technological objectives are outlined below:

\begin{itemize}
\item \textbf{Increase Network Observability:} RESISTO endeavors to revolutionize the design and monitoring of the electrical grid by enhancing its observability. This will be achieved through the installation of crucial sensors for predicting and detecting network faults, alongside the development of methodologies and software tools aimed at bolstering the capabilities of network control centers.
\item \textbf{Increase Situational Awareness:} The project aims to establish an infrastructure integrating comprehensive information about the study area. This includes the implementation of a network comprising sensors, autonomous aerial robots, and devices such as thermal cameras in distribution centers, facilitating tele-surveillance of overhead lines. Through this infrastructure, full IoT coverage will be achieved, enabling thorough monitoring and augmenting knowledge about the network's status.
\item \textbf{Increased Resilience:} Another primary objective of RESISTO is to enhance the resilience of the electrical network by detecting vulnerabilities and risks associated with extreme events in the long term. Corrective measures will be applied against identified vulnerabilities and risks, including fires and extreme winds. Continuous monitoring and short-term risk prediction will be conducted through integrated sensors, cameras, and software, enabling electrical control centers to anticipate disruptive events and activate emergency protocols promptly.
\item \textbf{Facilitate Operation and Maintenance (O\&M) Tasks:} Installation of sensors, thermal cameras, and remote surveillance systems will facilitate the development of algorithms for early detection of network failures. These algorithms will streamline O\&M tasks by enabling early anomaly detection in substations, lines, and distribution centers. Data obtained will support preventive maintenance of critical components such as transformers, mitigating risks of overheating and malfunction. Additionally, optimization of network operation during emergency situations will be facilitated.
\item \textbf{Adoption of Standardized Community Technologies:} RESISTO will adopt ICT standards developed by the European Union to enhance the existing European value chain, thereby promoting the adoption of standardized community technologies.
\item \textbf{Promote Conversion of Electrical Grid Sensor Network:} The project aims to leverage the potential of the implemented infrastructure to predict and monitor network risks. Furthermore, it will serve as a 'first responder' to competent authorities, offering valuable information for detecting risks affecting both nature and society.
\item \textbf{Increase Operational Capacity:} Through the project framework and installed technology, resources will be made available to network managers, significantly enhancing event detection in the electrical network. This heightened speed of action will lead to reduced affected areas, necessitating fewer resources for impact mitigation and ultimately increasing the operational capacity of competent agencies.
\end{itemize}

\section{Conclusions} The RESISTO project emerges as an innovative technological initiative and research transfer endeavor within Europe. In this article, the fundamental components of RESISTO are presented. First, the GridWatch electrical resilience platform serves as a comprehensive solution uniting climate data with the electrical transportation and distribution system. It aims to concentrate information for visualizing electrical system interactions, integrating external diagnostic tools, and utilizing artificial intelligence to detect and address electrical faults efficiently. The platform incorporates various tools and systems, including local weather stations, online weather sources, fire sensors, infrared cameras in transformation centers, and drones, to maximize information availability and facilitate informed decision-making. Particularly in this project, the GridWatch platform is integrated with a distributed network of 20 thermal cameras and servers for detecting thermal anomalies in power lines and electrical transformers, deployed in The Doñana National Park (Spain). This distributed network of cameras and servers features a machine learning-based system responsible for the recording, storage, analysis, and detection of thermal anomalies in different electrical installations. To accomplish this, it utilizes an adaptive algorithm that continuously learns from newly available temperature data and predicts the expected thermal behavior of the transformer accordingly. Upon detecting a thermal anomaly, it promptly notifies the GridWatch platform. Finally, in order to validate these thermal anomaly alarms, we have a fleet of drones consisting of DeltaQuad Pro for VTOL fixed-wing operations and DJI Matrice Series drones for multicopter operations equipped with thermal cameras. Additionally, a node has been developed to stream video from onboard cameras, enhancing real-time situational awareness during missions.

\begin{credits}
\subsubsection{\ackname} 
This work was supported by Spanish project RESISTO (2021/C005/00144188) funded by FEDER (Fondo Europeo de Desarrollo Regional) from Ministerio de Asuntos Económicos y Transformación Digital.
\subsubsection{\discintname}
\end{credits}
%
%
%
\bibliographystyle{unsrt}
\bibliography{samplepaper.bib}

\end{document}